\documentclass[pra,aps,amsmath,amssymb,twocolumn,showpacs]{revtex4-1}
\usepackage{graphicx}
\usepackage{color}
\usepackage{amssymb}
\usepackage{amsmath}
\DeclareMathSymbol{\lang}{\mathord}{symbols}{"68}
\DeclareMathSymbol{\rang}{\mathord}{symbols}{"69}
\DeclareMathSymbol{\openbra}{\mathord}{symbols}{"68}
\DeclareMathSymbol{\closeket}{\mathord}{symbols}{"69}
\DeclareMathOperator{\Rre}{Re}
\DeclareMathOperator{\Iim}{Im}

\newcommand{\ket}[1]{{| #1 \closeket}}

\newcommand{\aver}[1]{{\lang #1 \rang}}
\newcommand{\be}{\begin{equation}}
\newcommand{\ee}{\end{equation}}

\begin{document}
\title{Theory of dissipative chaotic atomic transport  in
an optical lattice}
\author{V.Yu. Argonov and S.V. Prants}
\affiliation{Laboratory of Nonlinear Dynamical Systems,\\
Pacific Oceanological Institute
of the Russian Academy of Sciences,\\ 43 Baltiiskaya st.,
690041 Vladivostok, Russia}
\begin{abstract}
We study dissipative transport of spontaneously emitting atoms in a 1D standing-wave
laser field in the regimes where the underlying deterministic Hamiltonian
dynamics is regular and chaotic. A Monte Carlo stochastic wavefunction method
is applied to simulate semiclassically the atomic
dynamics with coupled internal and translational degrees of
freedom. It is shown in numerical experiments and confirmed
 analytically that chaotic atomic transport can take the form either of ballistic motion or a
random walking with specific statistical properties. The character of spatial and momentum
diffusion in the ballistic atomic transport is shown to change abruptly in the atom-laser detuning
regime where the Hamiltonian dynamics is irregular in 
the sense of dynamical chaos. We find a clear correlation
between the behavior of the momentum diffusion coefficient
and Hamiltonian chaos probability which is a
manifestation of chaoticity of the fundamental
atom-light interaction in the diffusive-like dissipative
atomic transport. We propose to measure a linear extent of atomic
clouds in a 1D optical lattice and predict that, beginning with those values
of the mean cloud's momentum for which the probability of Hamiltonian chaos
is close to 1, the linear extent of the corresponding clouds should increase sharply.
A sensitive dependence of statistical characteristics of dissipative
transport on the values of the detuning allows to
manipulate the atomic transport by changing the laser frequency.
\end{abstract}
\maketitle

\section{Introduction}

An atom placed in a laser standing wave is acted upon by
two radiation forces, deterministic dipole and stochastic dissipative
ones \cite{Minogin,Kaz,Meystre}.
The mechanical action of light upon neutral atoms is at the
heart of laser cooling, trapping, and Bose-Einstein
condensation. Numerous applications of the mechanical action of light
include isotope separation, atomic litography and epitaxy,
atomic-beam deflection and splitting, manipulating
translational and internal atomic states, measurement of atomic positions, and
many others. Atoms and ions in an optical lattice, formed by a laser standing wave, are perspective objects for
implementation of quantum information processing and
quantum computing. Advances in
cooling and trapping of atoms, tailoring optical potentials
of a desired form and dimension (including one-dimensional
optical lattices), controlling the level of dissipation
and noise are now enabling the direct experiments
with single atoms to study fundamental 
 principles of quantum physics, quantum chaos, decoherence,
and quantum-classical correspondence (for recent reviews on cold atoms in optical lattices see Ref.~\cite{GR01,MO06}).

Experimental study of quantum chaos
has been carried out with ultracold atoms interacting in $\delta$-kicked
optical lattices \cite{MR94,RB95,DM95,Amm98,Rin00,From00,Hens03,JS04}.
To suppress spontaneous emission (SE) and provide a coherent quantum dynamics atoms in those experiments were 
 {\it detuned far
from the optical resonance}. Adiabatic elimination of
the excited state amplitude leads to an effective
Hamiltonian for the center-of-mass (CM) motion \cite{GSZ92},
whose 3/2 degree-of-freedom classical analogue
 has a mixed phase space with
regular islands embedded in a chaotic sea.
De Brogile waves of $\delta$-kicked ultracold atoms have been shown to
demonstrate under appropriate conditions the effect
of dynamical localization in momentum
distributions which means the quantum suppression of chaotic diffusion
\cite{MR94,RB95,DM95,Amm98,Rin00,From00,Hens03,JS04,GSZ92}. Decoherence
due to SE or noise tends to suppress this
 quantum effect and restore classical-like dynamics
 \cite{KO98,Ball99,SM00,Arcy01}. Another important 
 quantum chaotic
phenomenon with cold
 atoms in far detuned optical lattices
is a chaos assisted
tunneling. In experiments \cite{Steck01,HH01,Steck02} ultracold atoms have been
demonstrated to oscillate coherently
between two regular
regions in mixed phase space even though
the classical transport between
these regions is forbidden by a constant of motion (other than energy).

The transport of cold atoms in optical lattices
has been observed to take the form of ballistic motion, oscillations in wells of
the optical potential, Brownian motion \cite{Chu85}, anomalous diffusion and L\'evy
flights \cite{BB02,BB94,JD96,KS97,ME96}. The L\'evy flights
have been found in the context of subrecoil laser cooling \cite{BB02,BB94} in the distributions of
escape times for ultracold atoms trapped in the potential wells with
momentum states close to the dark state. In those experiments the variance and
 the mean time for atoms to leave the trap have been shown
to be infinite.

A new arena of
quantum nonlinear dynamics with atoms in optical
lattices is opened if we work {\it near the optical resonance} and take the
 dynamics of
internal atomic states into account. A single atom in a standing-wave
laser field may be treated as a nonlinear dynamical
system with coupled internal (electronical) and external (mechanical) degrees
of freedom \cite{PRA01,JETPL01,JETPL02}. In the semiclassical
and Hamiltonian limits (when one treats atoms as point-like particles and neglects
 SE and other losses of energy), a number of nonlinear dynamical
effects have been analytically and numerically demonstrated
with this system: chaotic Rabi oscillations 
 \cite{PRA01,JETPL01,JETPL02}, Hamiltonian chaotic atomic
transport and dynamical fractals \cite{JETP03,
 PLA03,pra07}, L\'evy flights and anomalous diffusion
 \cite{PRA02,JETPL02,JRLR06}. These effects are caused by
local instability of the CM motion in a laser
field. A set of atomic trajectories under certain conditions becomes exponentially sensitive to small
variations in initial quantum internal and classical external states
or/and in the control parameters, mainly, the atom-laser detuning. Hamiltonian evolution
is a smooth
process that is well described in a semiclassical approximation by the coupled Hamilton-Schr\"odinger
equations. A detailed theory of {\it Hamiltonian chaotic transport of
atoms} in a laser standing wave has been developed in our recent paper
\cite{pra07}.

The aim of the present paper is to study {\it dissipative
 chaotic transport of atoms} in a one-dimensional optical lattice
in the presence of SE events which
interrupt coherent Hamiltonian evolution at random
time instants. Generally speaking, deterministic (dynamical)
chaos is practically indistinguishable in some manifestations from a random (stochastic)
process. The problem becomes much more complicated when
noise acts on a dynamical system which is chaotic in the absence of noise. Such systems
are of a great practical interest. Comparatively weak noise may 
be treated as a small perturbation to deterministic equations
of motion, and one can study in which way the noise
 modifies deterministic evolution on different time
scales. However, SE is a specific shot quantum noise that cannot
be treated as a weak one because the internal state may
change significantly after the emission of a spontaneous photon. Special methods
are needed to describe correctly the dynamics of a spontaneously
emitting single atom in an optical lattice. The purpose of this
paper is twofold. Our first goal is to give a description
of possible regimes of dissipative atomic transport in the presence of SE and to
quantify their statistical properties. Our secondary intent
is a search for manifestations of the fundamental dynamical
instability and Hamiltonian atomic chaos in the
diffusive-like CM motion of spontaneously emitting atoms in a laser standing
wave which can be observed in real experiments.

The paper is organized as follows. In Sec. II we formulate
a Monte Carlo stochastic wavefunction approach to solving semiclassical
Hamilton-Schr\"odinger equations of motion for a two-level atom in a one-dimensional monochromatic standing light wave. This approach
allows to get the most probabilistic outcome that can be
compared directly with corresponding experimental
 output with single atoms. In Sec. III we review briefly
our previous results on Hamiltonian chaotic CM
motion which are necessary to quantify
and interpret dissipative dynamics.
Sec. IV is devoted to description of possible regimes of
dissipative CM motion of spontaneously emitting atoms in a standing
wave. Monte Carlo simulations of the well-known effects of
acceleration, deceleration, and velocity grouping,
and of a novel effect of {\it dissipative chaotic walking of atoms} are presented in this section. Anomalous statistical
properties of dissipative chaotic walking are quantified 
and discussed in Sec. V. Whereas Secs. IV and V are
devoted to solving the first task of this paper, in Sec. VI
we consider the problem of manifestations of dynamical
instability and Hamiltonian chaos in dissipative
atomic transport. We demonstrate analytically and numerically that character of diffusion of
spontaneously emitting atoms changes qualitatively
in the detuning regime where the underlying Hamiltonian dynamics
is chaotic. To observe this effect in a real experiment
 with cold atoms in a one-dimensional optical lattice we
propose to measure the linear extent of atomic clouds
with different values of their mean momentum
and predict that the extent should increase
significantly with those values of the mean momentum for
which the underlying Hamiltonian evolution is chaotic.

\section{Monte Carlo wavefunction modeling of the atomic dynamics}

In the
frame rotating with the laser frequency $\omega_f$, the standard Hamiltonian
of a two-level atom in a strong standing-wave 1D
laser field has the form
\begin{equation}
\begin{aligned}
\hat H=& \frac{\hat P^2}{2m_a}+\frac{1}{2}\hbar(\omega_a-\omega_f)\hat\sigma_z-
\\& -\hbar \Omega\left(\hat\sigma_-+\hat\sigma_+\right)\cos{k_f\hat X}-i\hbar\frac{\Gamma}{2}\hat\sigma_+\hat\sigma_-,
\label{Jaynes-Cum}
\end{aligned}
\end{equation}
where $\hat\sigma_{\pm, z}$ are the Pauli operators for the
internal atomic degrees of freedom, $\hat X$ and $\hat P$
are the atomic position and momentum operators,
$\omega_a$, $\omega_f$, and $\Omega$ are the atomic transition,
laser, and Rabi frequencies, respectively, and $\Gamma$
is the spontaneous decay rate. {Internal
atomic states are} described by the wavefunction
$\ket{\Psi(t)}=a(t)\ket{2}+b(t)\ket{1}$, with $a$ and $b$ being the
complex-valued probability amplitudes to find {an} 
atom in the excited $\ket{2}$ and ground $\ket{1}$
states. Note that the norm of the wavefunction,
$|a|^2+|b|^2$, is not conserved {due to non-Hermitean 
term in the Hamiltonian.}

 We use the standard Monte Carlo
wavefunction technique \cite{Carmichael}
to simulate the atomic dynamics with the coupled internal and
 external degrees of freedom in an optical lattice. 
 The evolution of an atomic state 
$\ket{\Psi(t)}$ consists of two parts: 
(i) jumps to the ground state ($a=0$, $|b|^2=1$) each of which is accompanied by the 
emission of an observable photon at random time moments 
with the mean time $2/\Gamma$ (actually, the probability of SE depends on the atomic population inversion) and (ii) coherent 
evolution with continuously decaying norm of the atomic state vector
without the emission of an observable photon. 
The decaying norm of the state vector is equal to the probability 
of spontaneous emission of the next photon.
 It is convenient to introduce
the new real-valued variables {(normalized all the time)} instead of the 
amplitudes $a$ and $b$ {(renormalized after SE events only)}
\begin{equation}
\begin{displaystyle}
u\equiv\frac{2\Rre\left(ab^*\right)}{\left|a\right|^2+\left|b\right|^2},
\quad
v\equiv\frac{-2\Iim\left(ab^*\right)}{\left|a\right|^2+\left|b\right|^2},
\quad
z\equiv\frac{\left|a\right|^2-\left|b\right|^2}{\left|a\right|^2+\left|b\right|^2},
\end{displaystyle}
\label{uvz_def}
\end{equation}
which have the meaning of synphase and quadrature
components of the atomic electric dipole moment
and the population inversion, respectively. We
stress that the length of the Bloch vector, $u^2+v^2+z^2=1$,
is conserved.

 Since we study manifestation of quantum nonlinear effects in ballistic transport
of atoms, when the average atomic momentum is very large as compared
with the photon momentum $\hbar k_f$, the translational motion is described classically by 
Hamilton equations. The whole atomic dynamics
is governed by the following Hamilton-Schr\"odinger
equations \cite{Acta06,epl08} 
\begin{equation}
\begin{aligned}
\dot x&=\omega_r p,\quad
\dot p=-u\sin x+\sum\limits_{j=1}^{\infty}\rho_j\delta(\tau-\tau_j),
\\
\dot u&=\Delta v+\frac{\gamma}{2}uz-u\sum\limits_{j=1}^{\infty}\delta(\tau-\tau_j),
\\
\dot v&=-\Delta u+2 z\cos x+\frac{\gamma}{2}vz-v\sum\limits_{j=1}^{\infty}\delta(\tau-\tau_j),
\\                                         
\dot z&=-2 v\cos x-\frac{\gamma}{2}(u^2+v^2)-(z+1)\sum\limits_{j=1}^{\infty}\delta(\tau-\tau_j),
\end{aligned}
\label{mainsys}
\end{equation}
where $x\equiv k_f \aver{\hat X}$ and $p\equiv \aver{\hat   P}/\hbar k_f$ are 
normalized atomic CM position and momentum.
The dot denotes differentiation with respect to
the normalized time $\tau\equiv \Omega t$. Throughout the paper we fix the values of the
normalized
decay rate $\gamma\equiv \Gamma/\Omega$ and the recoil frequency
$\omega_r\equiv\hbar k_f^2/m_a\Omega$ to be
$\gamma=3.3\cdot 10^{-3}$ and $\omega_r=10^{-5}$. This values are similar to those used in experiments
with Na \cite{MR94,RB95}, Cs \cite{Amm98,Hood00} and Rb \cite{Hens03} cold
atoms in a standing-wave laser field with the
maximal Rabi frequency of the order of $1\div 5$ GHz.
So, the normalized detuning between the field and atomic frequencies, $\Delta\equiv(\omega_f-\omega_a)/\Omega$, is a single variable parameter. Also we fix the initial conditions as follows: $x_0=v_0=u_0=0, z_0=-1$, and vary the initial momentum $p_0$ only. In Eqs. (\ref{mainsys}) $\tau_j$ are random
time moments of SE events
and $\rho_j$ are random recoil momenta with the
values between $\pm 1$ (1D case).
In terms of the normalized time $\tau$ the rate
 of occurrence of SE events is 
$\gamma (z+1)/2$. At  $\tau=\tau_j$, the
atomic variables change as follows:
\be 
\tau=\tau_j\Rightarrow u\to 0,\ v\to 0,\ z\to -1,\ p\to p+ \rho_j,\ -1\leq p_j\leq 1.
\label{puvz}
\ee
\section{A brief review of Hamiltonian atomic dynamics}

In this section we review briefly our main results on Hamiltonian
atomic dynamics (see Refs. \cite{PRA01,JETPL01,JETPL02,JETP03,JRLR06,pra07})
which will be used in the next sections. In the absence of any losses ($\gamma=0$) the total atomic energy is conserved,
   \begin{equation}
H\equiv\frac{\omega_r p^2}{2}+U,\quad U\equiv -u\cos x-\frac{\Delta}{2}z.
\label{H}
\end{equation}
The corresponding lossless equations of motion with two
independent integrals of motion, the energy $H$ and
the length of the Bloch vector, have been shown 
 \cite{JETPL01,PRA01} to be chaotic in the sense of an exponential
sensitivity to small variations in initial conditions and/or
the control parameters. The CM motion is governed
by the simple equation for a nonlinear physical pendulum
with the frequency modulation \cite{JETPLP02}
\be
\ddot x+\omega_r u(\tau)\sin x=0,
\label{pendulum}
\ee
where the synchronized component of the atomic dipole $u$ is
a function of all the other atomic variables including the translational ones.
Besides the regular CM motion, namely, oscillations in a well
of the optical potential and a ballistic motion over its
hills, we have found and quantified chaotic CM motion
\cite{JETPL01,PRA01,JETPLP02}. On the exact atom-laser resonance with
 $\Delta=0$, $u$ is a constant, and the CM performs either regular
oscillations, if $H< |u|$,
or moves ballistically, if $H>|u|$.

At $\Delta\neq 0$, the depth of the potential wells changes in
course of time, and atoms may wander in a rigid optical lattice
(without any modulations of its parameters) in a chaotic
way with alternating trappings in the wells and flights
of different lengths and directions over the hills. At small detunings,
$|\Delta|\ll 1$, the second term of the potential energy $U$ in Eq. (\ref{H}) is
small, and $U$ can be approximated by a function of only one internal variable
 $u$. In this case we have approximate solutions for $v$ and $z$
\begin{equation}
\begin{aligned}
v(\tau)=\pm\sqrt{1-u^2}\ \cos\left(2\int\limits_0^\tau \cos  x d\tau'+
\chi_0\right),\\
z(\tau)=\mp\sqrt{1-u^2}\ \sin\left(2\int\limits_0^\tau \cos  x d\tau'+
\chi_0\right),
\label{vz}
\end{aligned}
\end{equation}
where $\chi_0$ is an integration constant which is a function of initial values of $z$ and $u$.
Using these solutions one can prove that at $|\Delta|\ll 1$, $u$ performs shallow
oscillations when the atom moves between the nodes (recall that $u=const$ at $\Delta=0$).
These oscillations are synchronized with the oscillations of $z$, and when an atom approaches
any node with $\cos x=0$, where the  strength of the laser field changes the sign, they slow down (see Eq. (\ref{vz})).
 The swing of oscillations of $u$ gradually increases, 
and exactly at the node $u$
 changes abruptly its value (see Fig. \ref{fig1}).
Thus, $u$ is practically a constant between
the nodes and it performs a sudden jump at every
node. 

 In the Raman-Nath approximation, where
 $x=\omega_rp\tau$ and $p=const$, we have managed to derive the deterministic mapping
allowing to compute the value $u_m$ just after crossing the $m$th node
\begin{equation}
\begin{aligned}
u_m\simeq\sin\left\{\frac{\Delta}{\sqrt{1-u^2_{m-1}}} 
 \left[\sqrt\frac{\pi}{\omega_r p}\left(v_0\cos\left(\frac{2}{\omega_r p}-\frac{\pi}{4}\right)
+\right.\right.\right.\\\left.\left.\left.+(-1)^{m}z_0\sin\left(\frac{2}{\omega_r p}-\frac{\pi}{4}\right)\right)+(-1)^{m} z_0\right]+\arcsin u_{m-1}\right\},\label{udeterm} 
\end{aligned}
\ee
where $v_0$ and $z_0$ are the values of $v$ and $z$
at the antinodes of the standing wave at
 $x=\pi k$, $k=0,1,2,...$. They are the same at all
the antinodes because in the Raman-Nath approximation $v$ and $z$ are periodic functions of $x$ (see
 solution (\ref{vz})).
 Formula (\ref{udeterm}) describes the series of jumps of two alternating magnitudes (for odd and even $m$). Strictly speaking, (\ref{udeterm}) is valid with fast ballistic atoms
and not on a very long time scale. Deviation of the analytic
calculations with Eq.~(\ref{udeterm}) from the exact numerical results is demonstrated
in Fig.~\ref{fig1}a where we plot the function $u(\tau)$ for a fast atom with $p_0=1900$. It is obvious that the signal is rather
regular but the magnitude of the jumps changes slowly
because the Bloch components $v$ and $z$ are not strictly periodic
functions of time.

Figure \ref{fig1}b plots $u(\tau)$ in
the regime of Hamiltonian chaotic walking. To quantify chaotic
jumps of $u$ we proposed in Ref.~\cite{pra07} the stochastic
map
\begin{figure*}[htb]
\begin{center}
\includegraphics[width=0.8\textwidth,clip]{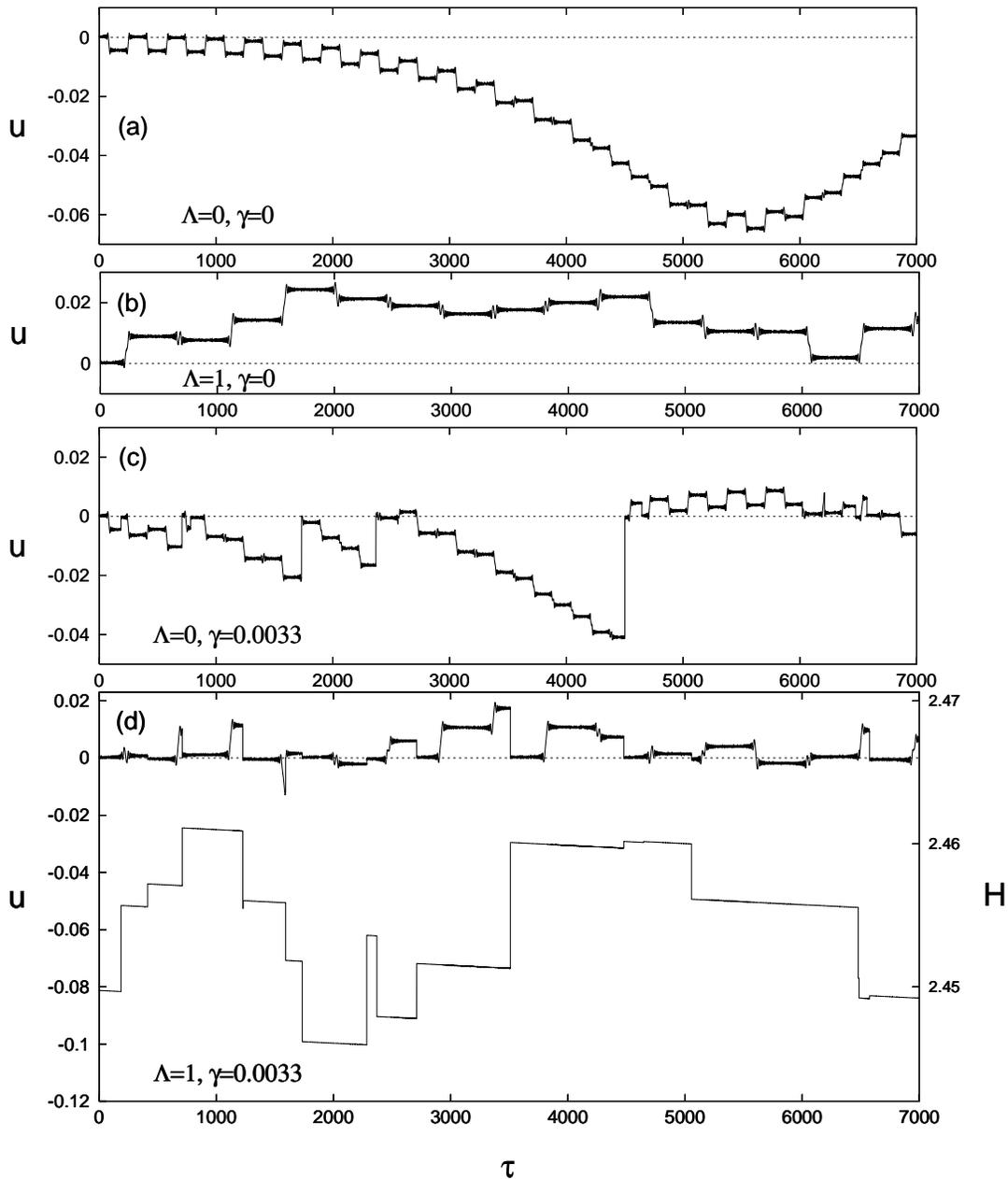}
\end{center}
\caption{Time evolution ($\tau$ is in units of $\Omega^{-1}$) of the synphase component of the electric dipole moment $u$
 and the atomic energy $H$ (in units of $\hbar\Omega$). (a) Regular Hamiltonian dynamics ($p_0=1900$, $\gamma=0$), (b) chaotic Hamiltonian dynamics
($p_0=700$, $\gamma=0$), (c) regular dissipative dynamics ($p_0=1900$, $\gamma=0.0033$), (d) chaotic dissipative dynamics ($p_0=700$,
 $\gamma=0.0033$). In all the panels, $\Delta=-0.0005$. The initial part of (a) agrees with approximate solution (\ref{udeterm}) with $v_0=0$, $z_0=-1$.}
\label{fig1}
\end{figure*}
\begin{equation}
\begin{aligned}
\ &u_m\simeq\sin\left(|\Delta|\sqrt\frac{\pi}{\omega_r p_{\rm{node}}}
\sin\phi_{m}+\arcsin u_{m-1}\right),
\end{aligned}
\label{u_m}
\end{equation}
which was derived from the deterministic map (\ref{udeterm}) by introducing
random phases $\phi_m$ $(0\leq\phi\leq 2\pi)$ instead of arguments of trigonometric functions
which may differ significantly from node to node due to strong variations in the
atomic momentum beyond the Raman-Nath approximation. Note that the value of the momentum at the instant
when the atom crosses a node, $p_{\rm node}=\sqrt{2H/\omega_r}$, is
approximately the same for all nodes.

The map (\ref{u_m}) describes a random Markov
process in the $u$ space with $u_m$ varying in the range $-1\leq u_m\leq 1$. This quantity may be
smaller or larger than the atomic energy $H$ (which is a constant in the
Hamiltonian limit). Since the values of $u$ define the atomic potential energy,
 its random walking governs a random walking of atoms in the lattice.
The possible regimes of the Hamiltonian CM motion can be summarized as follows \cite{pra07}:
At $|u|>H$, an atom oscillates in one of the potential wells, at $|u|<H$,
it moves ballistically. It can walk chaotically if $0<H<1$.
In the process of Hamiltonian chaotic walking the atom wanders
in an optical lattice with alternating trappings in wells of the optical
potential and flights over its hills changing the direction of motion many times. ``A flight'' is an event
when the atom passes, at least, three
 nodes. CM oscillations in a well of the optical
potential is called ``a trapping''. The number of node
crossings $l$ during a single flight or a single trapping event is a discrete
measure of the length and durations of those events. We have derived in Ref.~\cite{pra07}
the following formulas for the probability density functions (PDFs) for
the flight and trapping events in the diffusive approximation:
\be
\begin{aligned}
P_{fl}(l )\simeq\frac{Q(D_u)}{\arcsin^3 H}&\sum\limits_{j=0}^{\infty}(j+1/2)^2*\\
&*\exp\frac{-(j+1/2)^2\pi^2 D_u l }{\arcsin^2 H},\\
P_{tr}(l )\simeq\frac{Q(D_u)}{\arccos^3 H}&\sum\limits_{j=0}^{\infty}(j+1/2)^2*\\
&*\exp\frac{-(j+1/2)^2\pi^2 D_u l }{\arccos^2 H}.
\label{stat}
\end{aligned}
\ee
Here $Q$ is a constant, $D_u=\Delta^2\pi/4\omega_r p_{\rm{node}}$
is a diffusion coefficient in the $u$ space. For comparatively small values of $l$
(i. e., with short flights and trappings), we get from Eq.~(\ref{stat})
the power decay
\begin{equation}
P_{fl}\propto P_{tr}\propto l^{-1.5},
\end{equation}
whereas for large $l$ the decay is exponential. Numerical simulation of the
Hamiltonian equations of motion agrees well with the
analytical results (\ref{stat}) in different ranges of the detunings. A typical PDF for
the flight and trapping events decays initially algebraically
 and has an exponential tail. The length of the initial power-law segment
is inversely proportional to the 
value of the detuning $\Delta$ and can be rather large.

In which way SE changes the statistical properties of the Hamiltonian motion?
 Can we find fingerprints of Hamiltonian instability
and chaos in the motion of spontaneously emitting atoms
or SE totally suppresses any manifestations of coherent (but chaotic!) Hamiltonian
 dynamics? These questions will be addressed in the next sections.

\section{Dissipative atomic transport in a laser standing wave}

The emission of a photon into the continuum of modes of the
electromagnetic field is accompanied by an atomic recoil.
The dissipative (friction) force $F\equiv\aver{\dot p}$ (which does not exist in the Hamiltonian system) depends on the atomic momentum $p$ and the sign of the detuning
in a complicated way \cite{Kaz,PRA05}.
 The effects of
acceleration, deceleration, and velocity
grouping (at $\Delta<0$) are well-known in the
literature \cite{Kaz,Meystre}. A novel effect we report in this section is
{\it dissipative chaotic walking}. It appears under
appropriate conditions that are different from those specified
for {\it Hamiltonian chaotic walking} in the preceding
section.

To illustrate the possible regimes of dissipative
 atomic transport in a standing wave we integrate by
the Monte Carlo method dissipative equations of motion
(\ref{mainsys}) with 2000 atoms whose positions and
momenta are distributed in a quasi-Gaussian manner
(Fig.~\ref{fig2}a). In Fig.~\ref{fig2}b we demonstrate the
\begin{figure*}[htb]
\begin{center}
\includegraphics[width=0.9\textwidth,clip]{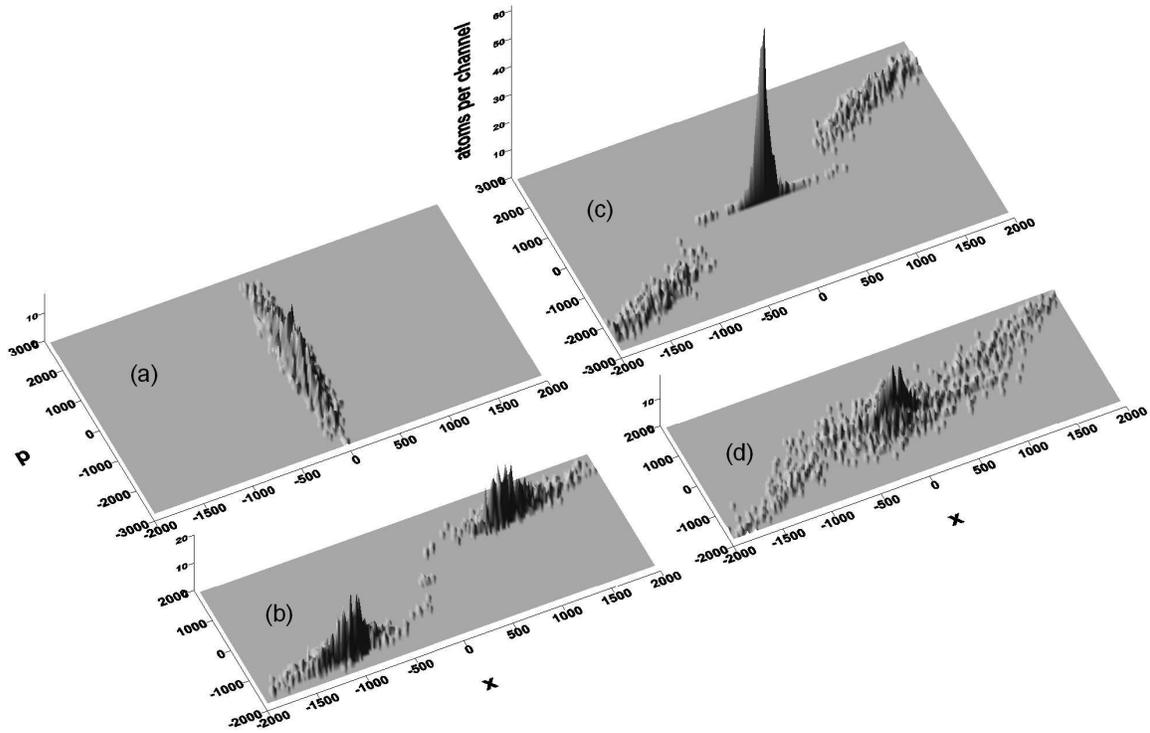}
\caption{Atomic momentum and position distribution illustrating the effects of atomic acceleration, deceleration,
and the velocity grouping: (a) $\tau=0$,
(b) $\tau=10^{5}$, $\Delta=-0.2$, (c) $\tau=10^{5}$, $\Delta=0.1$,
(d) $\tau=10^{5}$, $\Delta=-0.05$. Momentum $p$ is given in units of
$\hbar k_f$, the position in units of $k_f^{-1}$.}
\label{fig2}
\end{center}
\end{figure*}
velocity grouping effect at $\Delta=-0.2$ and $\tau=10^5$.
A large number of atoms is grouped around two
 values of the capture momentum $p_{\rm cap}\simeq\pm 1300$ because of acceleration of slow atoms
 and deceleration of the fast ones in the initial ensemble.
The slower the atoms are the longer is the process
 of the velocity grouping. Note that atoms with $|p|\lesssim 100$,
trapped initially in a well of the optical potential, 
could not quit the well up to $\tau=10^5$.
Contrary to that, at positive values of the detuning
fast atoms are accelerated and slow ones are decelerated.
As a result, we observe a pronounced peak around $x\simeq p\simeq 0$
shown in Fig.~\ref{fig2}c at $\Delta=0.1$ and $\tau=10^5$.

Dependence of the friction force $F$ on the current atomic
momentum $p$ is shown in Fig.~\ref{fig3} at $\Delta=-0.2$.
It has been computed with our main equations (\ref{mainsys})
when averaging over seven thousands atoms with
different initial momentum. The function $F(p)$ resembles
the behavior of the friction force computed with another methods
\begin{figure}
\begin{center}
\includegraphics[width=0.49\textwidth]{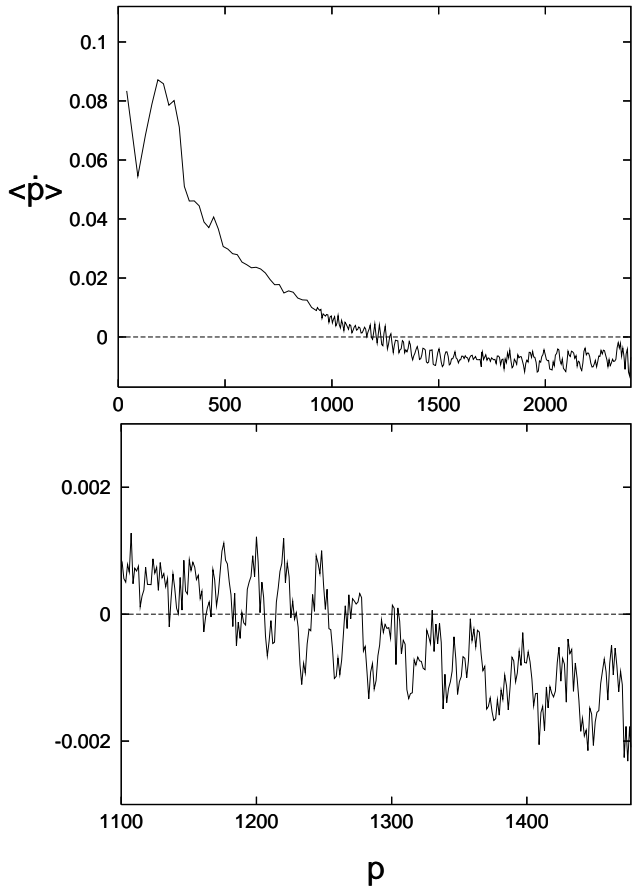}
\caption{Dependence of the friction force $F\equiv\aver{\dot p}$ on the current atomic momentum $p$ at the detuning $\Delta=-0.2$.}
\label{fig3}
\end{center}
\end{figure}
 (see \cite{Kaz} and Fig. 1a
in \cite{PRA05}). The friction force decreases up to
zero value and then begin to oscillate with
increasing $p$. It has a number of zeroes (the detailed view is shown in Fig.~\ref{fig3}b) like
the corresponding functions in Refs.~\cite{Kaz,PRA05}.
Zero values of $F$ correspond to quasistationary values of the momentum which depend on
$\Delta$. Some of them are attractors and atoms with close
values of the momentum tend to $p_{\rm cap}$, another ones
are repellors. The attractors and repellors are not
deterministic because of a random nature of SE.
Thus, an atoms walks randomly in the momentum
space between different values of the capture
 momentum $p_{\rm cap}$. When it reaches a certain value of
 the capture momentum the atom does not stop in the momentum space
and goes on to fluctuate because of both
the Hamiltonian instability and SE effect.

In the preceding section we described the Hamiltonian
chaotic walking that may occur in the absence of
any losses.
Dissipation causes additional strong
fluctuations of the momentum. If $\Delta>0$ or if it is negative but comparatively large, nothing principally
new happens to atoms as compared with the Hamiltonian limit. However, at
negative small values of $\Delta$, a characteristic capture momentum becomes smaller
than a typical range of momentum fluctuations due to
atomic recoils. As a result, atoms may change their
direction of motion in an irregular way. Such a {\it
dissipative chaotic atomic walking} is illustrated in
Fig.~\ref{fig2}d at $\Delta=-0.05$ and $\tau=10^5$ with the atoms distributed widely in the phase plane. Typical atomic trajectories
 are shown in Fig.~\ref{fig4} in the momentum space.
Figures \ref{fig4}a and b illustrate how the friction force near the resonance ($\Delta=-0.001$ and $\Delta=-0.01$)
decelerates atoms with large values of
the initial momentum down to so small values of the capture momentum when the dissipative
chaotic walking becomes possible. With increasing the absolute
value of the negative $\Delta$, the capture momentum increases and
the atom changes rarely the direction of motion (Fig.~\ref{fig4}c
with $\Delta=-0.1$). Panels d and e in Fig.~\ref{fig4} illustrate the
velocity grouping effect at $\Delta=-0.15$ with different values of the initial momentum.

\begin{figure}
\includegraphics[width=0.49\textwidth]{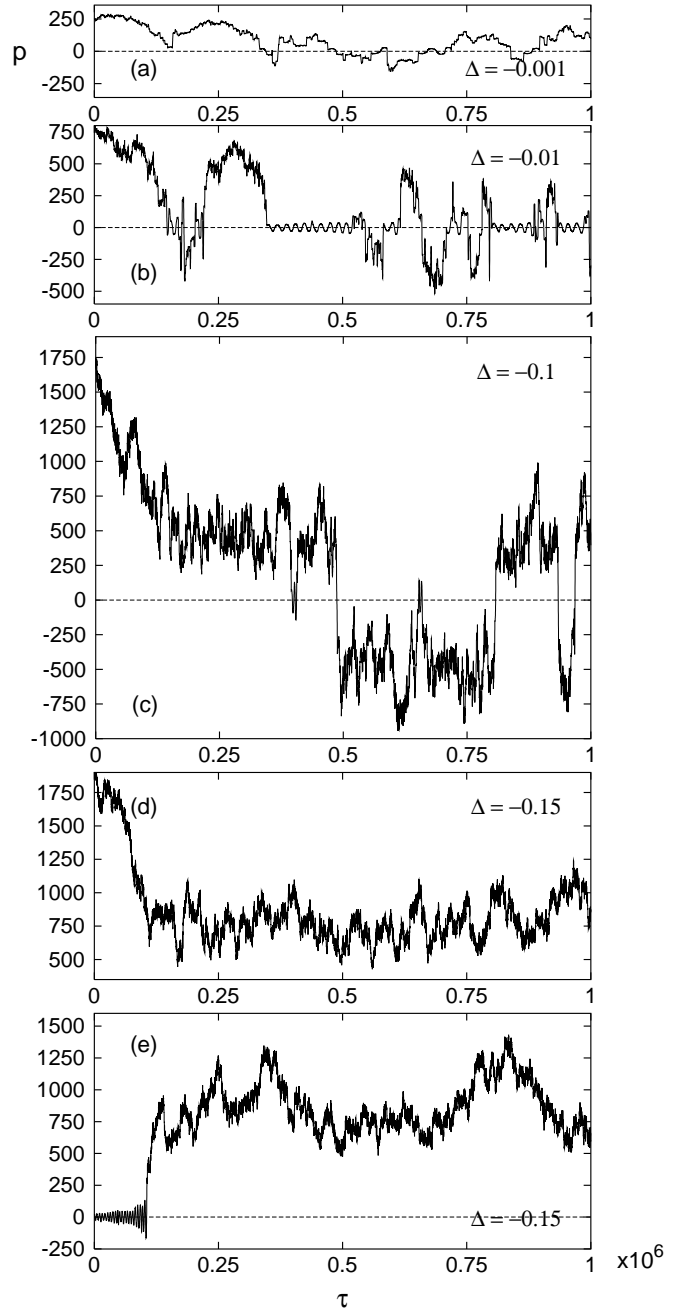}
\caption{Typical atomic trajectories in the momentum space: (a)-(c) dissipative chaotic walking with
different statistics of atomic flights, (d)-(e) the effect of velocity grouping. Note that atoms
with very different initial momentum acquire a close value of the capture momentum.}
\label{fig4}
\end{figure}

\section{Statistical properties of dissipative chaotic walking}

Statistics of Hamiltonian chaotic walking is quantified by the
flight and trapping PDFs (\ref{stat}) with algebraically decaying
head segments and exponential tails whose lengths strongly depend
on $\Delta$. We will show in this section that PDFs for dissipative chaotic walking are even more sensitive to variations in $\Delta$.
Figures \ref{fig4}a and b clearly demonstrate that at very small detuning $\Delta=-0.001$ long flights dominate,
whereas there appears a number of
short flights with larger value of $\Delta=-0.01$. In Fig.~\ref{fig5} we plot the PDFs $P_{fl}$
for the duration of atomic flights $T$ with
different values of the detuning $\Delta$. At small detunings (Fig.~\ref{fig5}a),
the length of the power-law segments depends on $\Delta$ in a similar way
as in the Hamiltonian case (compare this figure with Figs. 5a, 6a, and 7a in Ref.~\cite{pra07}).
 However, the slope slightly differs from the Hamiltonian slope which is equal to $-1.5$. The difference
in the statistics of dissipative and Hamiltonian walkings is more evident
with larger values of the detuning (Fig.~\ref{fig5}b). The length of the power-law segments increases
 drastically with increasing $\Delta$. This effect is absent in Hamiltonian dynamics. The corresponding slope $\alpha$ decreases with changing
the detuning from $\Delta=-0.09$ to $\Delta=-0.12$ because of the corresponding increase
in the length of atomic flights (see Figs.~\ref{fig4}a, b, and c). In Fig.~\ref{fig6} we plot the dependencies of the mean
duration of atomic flights $\aver{T}$ and
the slope of the PDF powerlaw fragments $\alpha$ on the detuning
$\Delta$ in the range of its medium values $-0.12\leq\Delta\leq-0.06$.
Both the quantities correlate well with each other. It means that, changing
 the value of $\Delta$, one can manipulate statistical properties
of dissipative atomic transport in an optical
lattice. This control is nonlinear in the sense
that slightly changing $\Delta$, say, from $-0.08$ to $-0.12$,
we increase the mean duration of flights in a
few orders of magnitude. This effect may be qualitatively explained
as follows. When increasing the absolute value of the negative detuning, the capture momentum increases
but fluctuations of the current momentum $p$ decrease providing long atomic flights \cite{epl08}.

To explain the statistical properties of the dissipative chaotic walking
let us consider the behavior of the quasienergy
\begin{figure}
\includegraphics{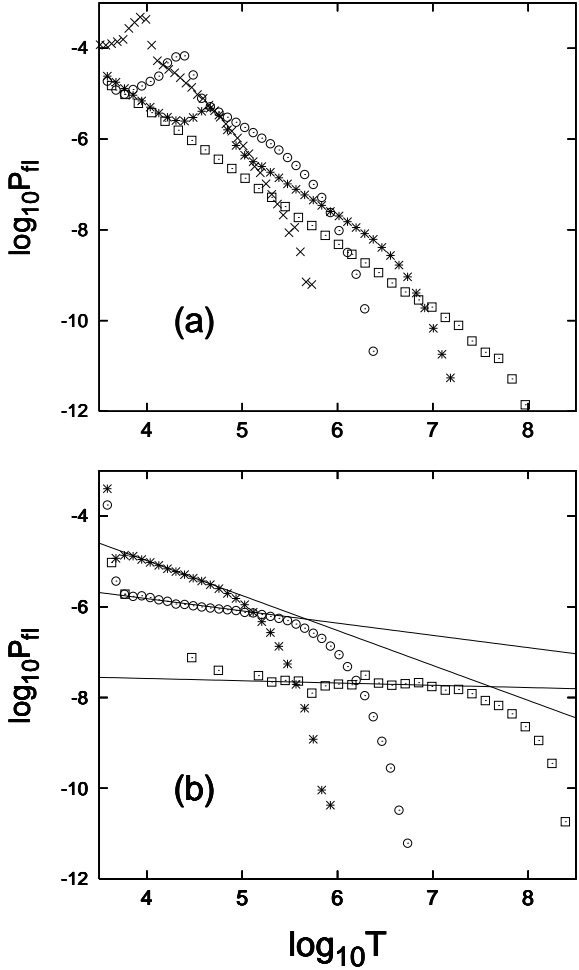}
\caption{The PDFs $P_{\rm fl}$ for the duration of atomic flights $T$ with 
(a) small detunings (crosses $\Delta=-0.01$, stars $\Delta=-0.001$, 
circles $\Delta=-0.0001$, squares $\Delta=-0.00001$) and 
(b) medium detunings (stars $\Delta=-0.09$, $\alpha = -0.77$; 
circles $\Delta=-0.1$, $\alpha = -0.27$; squares $\Delta=-0.12$, 
$\alpha = -0.05$). Straight lines show slopes $\alpha$
of the power-law fragments of the PDFs in log-log scale. }
\label{fig5}
\end{figure}
\begin{figure}
\includegraphics[width=0.49\textwidth,clip]{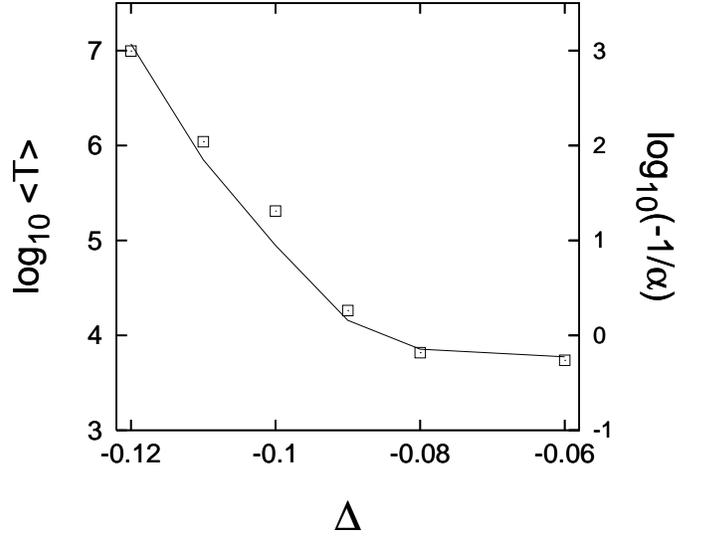}
\caption{Dependencies of the logarithms of the mean duration of
atomic flights $\langle T\rangle$ (solid line)
and of the slope $\alpha$ of the PDF power-law fragments (squares)
on the detuning $\Delta$ (in units of $\Omega$).}
\label{fig6}
\end{figure}
\begin{equation}
\begin{aligned}
\tilde H_j &\equiv\frac{\omega_r}{2}p^2-u\cos x-\frac{\Delta}{2}z-\frac{\Delta\gamma}{4}\aver{1-z^2}(\tau-\tau_j)=\\
&=H-\frac{\Delta\gamma}{4}\aver{1-z^2}(\tau-\tau_j),\quad
\tau_j<\tau<\tau_{j-1},
\end{aligned}
\label{PH}
\end{equation}
which is equal to the total atomic energy (\ref{H}) in
the absence of relaxation. Near the resonance, $|\Delta|\ll 1$, $\tilde H_j$ is
almost conserved between SE events, i. e., in the interval $\tau_j<\tau<\tau_{j-1}$. The real energy $H$
(see Fig.~\ref{fig1}d) 
decreases a little in between in a linear way. The rate of this decrease
is defined by the coefficients of spontaneous emission $\gamma$,
 the detuning $\Delta$, and the average probability to find the atom
in the excited state.
Both the quantities, $H$ and $\tilde H$, changes abruptly just after SE (because
of the corresponding changes in the atomic variables (\ref{puvz})). Just after emitting a $j$th spontaneous photon at $\tau=\tau_j$, they have the same values.
 So, we will model
the evolution of the energy as a map $H_j\equiv H(\tau_j^+)$ taken at the
moments $\tau_j^+$ just after SE
\be
\begin{aligned}
&H_j=\tilde H_j-\tilde H_{j-1}+H_{j-1}=H_{j-1}+\omega_r p(\tau_j^-) \rho_j+\\&+\frac{\omega_r}{2} \rho_j^2+
 \frac{\Delta}{2}+u(\tau_j^-)\cos x(\tau_j)+\\&+\frac{\Delta}{2}z(\tau_j^-)+\frac{\Delta\gamma}{4}\aver{1-z^2}(\tau_j-\tau_{j-1}),
\label{DH}
\end{aligned}
\end{equation}
where the values of the atomic variables $p(\tau_j^-),u(\tau_j^-)$, and $z(\tau_j^-   )$
are taken at the moments $\tau_j^-$ just before SE. They are, in turn, determined by the coherent evolution between SE.

The stochastic map for the atomic energy (\ref{DH})
provides an important information about the CM motion.
It has been shown in Ref.~\cite{pra07} that atoms
move ballistically, if the atomic energy satisfies to the condition
 $H\gtrsim |u|$, whereas at $H\lesssim |u|$
 they may change the direction of motion. The dissipative
chaotic walking takes place when the atomic energy $H$
alternatively takes the values larger and smaller than
a critical value $H=|u|$. In the Hamiltonian
limit, where the energy is conserved, the problem of
the CM chaotic walking has been reduced to the task of
random walking of the Bloch component $u$ (see Sec. III and Ref.~\cite{pra07}).
The energy is not conserved in the presence of relaxation, but the values of $u$ are
always small (see Appendix and Fig.~\ref{fig1}c and d). Thus,
atoms oscillate in the wells of the optical
potential if $H\lesssim 0$ and move ballistically
if $H\gtrsim0$.

On a time scale exceeding the mean time between SE events $2/\gamma$, the
evolution of energy can be treated as a diffusion process with a drift in the energetic space.
The probability to have the energy $H$
 at time $\tau$ is governed by the Fokker-Planck equation
\be
\dot P(H, \tau)=-2c_H\frac{\partial P}{\partial H}+ D_H\frac{\partial^2 P}{\partial H^2},
\label{fok}
\ee
where $D_H$ is an energy diffusion coefficient and $c_H$ is an energy drift coefficient which
can be estimated with the help of Eq. (\ref{DH}) as follows:
\be
c_H\equiv\frac{\aver{H_j- H_{j-1}}}{\aver{\tau_j-\tau_{j-}}}=\aver{\dot H}\simeq\frac{\omega_r\gamma}{12}+\frac{\Delta\gamma}{2}.
\label{dotH}\end{equation}
In deriving this formula we adopt the average value of the squared
recoil momentum $\aver{\rho_j^2}=1/3$ (the projections of the recoil momenta on the standing-wave axis $x$, $\rho_j$, are assumed to be distributed in the range $\pm 1$ with the same probability), the
average value of the population inversion just before a
 SE event to be $z(\tau_j^-   )= 1/2$, the average value of $z$ over the whole time scale and its mean squared deviation from 1 to be $\aver{z}=0$ and $\aver{1-z^2}=1/2$, respectively (see the solution (\ref{vz2}) in Appendix).
 Moreover, neglecting the correlation, we put $\aver{u\cos x}\simeq\aver{u}\aver{\cos x}=0 $,
which is valid if $|\omega_r p|\gtrsim\gamma/2$, i. e., when $p\gtrsim100$ with our choice of the parameters. Since the first term in (\ref{dotH})
is small and may be neglected, the drift velocity of an atom in the energetic
space is approximately proportional to the detuning $\Delta$, and, therefore
 in average atoms accelerate and decelerate at $\Delta>0$ and
$\Delta<0$, respectively, as it should be for $|\Delta|\ll 1$. In the weak Raman-Nath approximation,
 (\ref{Hsim}) and (\ref{ptauj}), the drift coefficient in the energetic
space is simply related to the friction force $F$ acting upon atoms
\be
F\equiv\aver{\dot p}\simeq\frac{\aver{\dot H}}{\omega_r p}.
\ee
The friction force plays the role of a drift coefficient in the
momentum space. Strictly speaking, the weak Raman-Nath approximation is not
valid near the turning points  when the atomic velocity is comparatively small.
However, most of the flight time it is valid.

The diffusion coefficient in the energetic space
is given by the formula
\be
\begin{aligned}
\ D_H\equiv\frac{\aver{(H_j-H_{j-1})^2}-\aver{H_j-H_{j-1}}^2}{2\aver{\tau_j
-\tau_{j-1}}},
\label{ddef}
\end{aligned}
\ee
which can be rewritten with the help of
 (\ref{DH}) as follows:
\be
\begin{aligned}
D_H\simeq\frac{\gamma\omega_r^2 p^2(\tau_j^-)}{{12}}+\frac{\aver{u^2(\tau_j^-)}\gamma}{8}.
\label{dh2}
 \end{aligned}
\ee
Using weak Raman-Nath approximation, (\ref{Hsim}) and (\ref{ptauj}), the first term can
be replaced by $\gamma\omega_rH/6$. Using the estimation (\ref{u2_m2}) for $\aver{u^2(\tau_j^-)}$
in the irregular CM motion regime
 (see Appendix), we get the following expression for the energy diffusion coefficient:
\be
D_H^{ch}\simeq\frac{\gamma\omega_r H}{{6}}+\frac{\Delta^2}{8}.
\label{dhc}
\ee
This expression is valid for moderately small momentums ($p\lesssim 1000$) when
the strong Raman-Nath approximation cannot be applied. In the process of dissipative chaotic walking,
the probability to get higher values of the momentums is almost zero.

Now we will try to derive analytically a distribution of the durations
$T$ of atomic flights in the process of dissipative
chaotic walking. In fact, it is a problem of the
first passage time for the atomic energy $H$ to return to its
zero value. Recall that at small detunings we have $H\simeq0$
in the very beginning of every flight. In course of time
$H$ can reach rather large values, and it returns to zero at the end
of the flight. If the random
 jumps of the energy would be symmetric ($c_H=0$), the probability to
find a flight with duration $T$ would be
proportional to $T^{-1.5}$, where the exponent $-1.5$
does not depend on the diffusion coefficient. This
conjecture follows from the known theorem in
 probability theory. More general result (see chapter XIV in Ref.~\cite{Feller}) proves that in the case
of an asymmetric random walking in the energetic space ($c_H\neq 0$) the PDF for
the flight durations in configuration space is 
\begin{equation}
P_{\rm{fl}}\propto e^{-c_H^2T/ D_H}T^{-1.5},
\label{statdissipa}
\end{equation}
if the drift and diffusion coefficients in the Fokker-Planck
 equation for the random walking are assumed
to be constants. This formula gives a distribution
of the flight durations with a power-law fragment
followed by an exponential tail and agrees qualitatively
with the exact numerical computations of
$P_{fl}$ shown in Fig.~\ref{fig5}a for a few values of
 the detuning $\Delta$. The main disadvantage of this formula is that (\ref{statdissipa}) does not depend on $\Delta$ as the exact PDFs in Fig.~\ref{fig5}a.
At very small $\Delta=-10^{-5}$, the PDF, shown by squares in
 Fig.~\ref{fig5}a, decays mostly algebraically,
 whereas at larger values of the detuning the
power-law fragments are much shorter.
 A discrepancy between the analytical and numerical PDFs arises
because we assumed in deriving (\ref{statdissipa}) that
 $D_H$ and $c_H$ do not depend on the energy $H$. In fact, it is
 not the case for small values of the momentum $p$, and a more accurate formula for $P_{fl}(T)$ is required.

The PDFs for Hamiltonian (\ref{stat}) and dissipative (\ref{statdissipa})
transport are similar in the sense that both $P_{fl}$ contain power-law
fragments followed by exponential tails, but the origin of each statistics is
different. In the Hamiltonian limit the statistics is governed by the
behavior of $u$, not the energy, as in the dissipative system. A
turnover from a power law to an exponential decay in the Hamiltonian
 case is explained by a boundedness of $|u|\lesssim 1$, whereas in the
dissipative system it is explained by a negative drift of the energy $H$. Each of the
factors prevents the corresponding randomly walking quantity to go
far away from its critical value (at which the atoms can change
the directions of motion) decreasing the probability of long
flights in at exponential way.

\section{Manifestation of Hamiltonian chaos in dissipative atomic transport}

In the absence of SE the atomic dynamics can be regular or chaotic depending
 on the initial conditions and/or the detuning. In experiments one measures statistical
 characteristics of spontaneously emitting atoms. Is there a correlation between those characteristics
and the underlying Hamiltonian dynamics? Can we find any
 manifestations of Hamiltonian instability, chaos, and order, in the diffusive-like dissipative atomic transport?
These questions will be addressed in the present section.

The common quantitative criterion of deterministic chaos, the maximal
Lyapunov exponent $\lambda$, is a measure of a divergence of
two trajectories in the phase space with close initial conditions
\cite{LL83}. To quantify probability of chaos in
 the mixed Hamiltonian dynamics, when $\lambda=0$ with some values of $p_0$
and $\lambda>0$ with another values of $p_0$,
 we introduce a probabilistic measure of
 Hamiltonian chaos
\be
\Lambda\equiv\aver{2\Theta(\lambda)-1}, 
\ee
where $\Theta(\lambda)$ is a Heaviside function ($\Theta=0$ if $\lambda<0$,
$\Theta=1/2$ if $\lambda=0$, and $\Theta=1$ if $\lambda>0$). {\it The probability
of Hamiltonian chaos} $\Lambda$ is computed by averaging over
a large number of atomic trajectories with different values
of $p_0$. If all the trajectories in the set turn out to be stable,
one gets $\Lambda=0$, and if all they are exponentially unstable, then
 $\Lambda=1$. One gets $0<\Lambda<1$, if some trajectories in the set are
 stable but the other ones are not. The magnitude of
$\Lambda$ is proportional to the fraction of trajectories with 
positive $\lambda$s.

To examine manifestations of the underlying Hamiltonian dynamics in
dissipative transport in is convenient to consider
atomic diffusion not in the energetic but in the momentum
 space. The momentum diffusion coefficient, which is a measure
of momentum fluctuations, can be written with the help of (\ref{ddef}) and (\ref{Hsim})
as follows:
\begin{figure*}[htb]
\includegraphics[width=0.96\textwidth,clip]{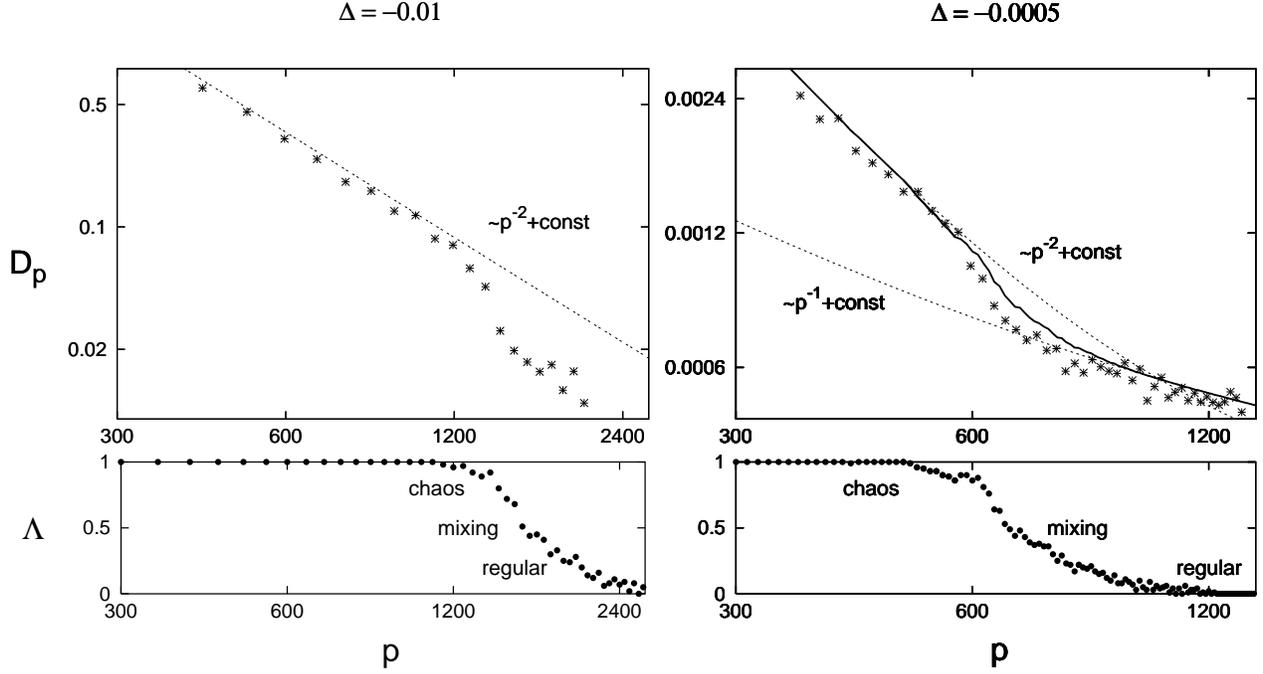}
\caption{Correlation between the average momentum diffusion coefficient 
in a log-log scale $D_p$ (in units of $\hbar^2k^2_f\Omega$) and probability of 
Hamiltonian chaos $\Lambda$ in
their dependencies on current atomic momentum $p$ (in units of $\hbar k_f$) 
at $\Delta=-0.01$ and $\Delta=-0.0005$.
Dashed lines with the slopes $~p^{-2} + {\rm const}$ and $~p^{-1} + {\rm const}$ 
are theoretical curves (\ref{dpc}) and (\ref{dpreg}) valid in the regimes of 
Hamiltonian chaos and order, respectively. Solid line is a theoretical curve 
(\ref{dprdpc}) derived to fit the exact numerical results. An abrupt change in the decay laws for 
$D_p$  occurs for those values of $p$ where the transition from order to 
chaos takes place in the underlying Hamiltonian dynamics.}
\label{fig7}
\end{figure*}
\be
\begin{aligned}
D_p\simeq\frac{D_H}{\omega_r^2p^2}\simeq\frac{\gamma}{{12}}+\frac{\aver{u^2(\tau_j^-)}\gamma}{8\omega_r^2p^2}.
\label{D2}
 \end{aligned}
\ee
Using the formula (\ref{dhc}), we get $D_p$ in the regime of chaotic oscillations of the Bloch
component $u$
\be
D_p^{ch}\simeq\frac{\gamma}{{12}}+\frac{\Delta^2}{8\omega_r^2p^2}.
\label{dpc}
\ee
The momentum diffusion coefficient $D_p$ is computed
with the main equations (\ref{mainsys}) in the following way.
The range of all possible values of the atomic energy $H$ (\ref{H})
is partitioned in a large number of bins. For a large
number of initial conditions (in fact, we change only the
 initial momentum $p_0$ keeping the other conditions to
be fixed), after any $j$th SE event we compute the difference $H_j-H_{j-1}$ and the squared difference $(H_j-H_{j-1})^2$.
 They are random values, but their statistics depend on
 the preceding energy value $H_{j-1}$. So we calculate the histograms of the
average values of $\aver{H_j-H_{j-1}}$ and $\aver{(H_j-H_{j-1})^2}$ as
 functions of energy $H_{j-1}$.
 After that we can compute the energy diffusion coefficient $D_H$
 (\ref{ddef}) which, being divided by $\omega_r^2p^2$, yields the
momentum diffusion coefficient $D_p$ which is better to present as a
 function of the current momentum $p\simeq\sqrt{2H_{j-1}/\omega_r}$.

The main result in this section is illustrated with Fig.~\ref{fig7}. In the
 upper left and right panels
the dependencies of the momentum diffusion coefficient $D_p$ on the current
momentum $p$ are plotted in a log-log scale for $\Delta=-0.01$ and $\Delta=-0.0005$,
respectively. In both the cases, we put $\gamma=0.0033$. These plots should be compared with the
corresponding lower panels where the probability of the Hamiltonian chaos $\Lambda$
is plotted against $p$ with $\gamma=0$ (i. e. in the Hamiltonian limit of Eqs. (\ref{mainsys})).
It is evident that the
character of the momentum diffusion changes abruptly
at those values of the current momentum where a transition from chaos
to order occurs in the underlying Hamiltonian dynamics. Such a
 turnover takes place
 in a range of small negative detunings and is a
manifestation of the peculiarities of the underlying Hamiltonian
evolution in the diffusive-like dissipative transport of atoms
in a standing-wave laser field. We may conclude that
in spite of random atomic recoils due to SE the chaotic
(regular) dynamics between the acts of SE
clearly manifests itself in the behavior of the measurable characteristic of the
 atomic transport, the momentum diffusion coefficient $D_p$.
The behavior of $D_p$ in the range of $p$, where the underlying
Hamiltonian evolution is chaotic, is well described by
the formula (\ref{dpc}) with $D_p^{ch}\sim p^2+const$ (see both
the upper panels in Fig.~\ref{fig7} where this dependence is
shown by dashed lines).

However, the formula (\ref{dpc}) does not work in the regimes
when the underlying Hamiltonian dynamics is mixed or regular because in deriving it we supposed
fully chaotic behavior of $u$. We have managed to estimate analytically $D_p$
in the Hamiltonian regular regime at extremely small values of the detuning $|\Delta|\lll 1$
and for atoms whose momentum is so large that we can neglect
its fluctuations between SE events (the exact Raman-Nath approximation with $x=\omega_rp\tau$).
Figure \ref{fig1}a illustrates the ladder-like behavior of $u$ which
 is descibed by the deterministic mapping (\ref{udeterm}) on a comparatively short timescale. To get $D_p^{RN}$ from Eq. (\ref{D2})
 we use the expression (\ref{u2determ2}) for $u^2(\tau_j^-)$ derived in Appendix
\be
D_p^{RN}\simeq\frac{\gamma}{{12}}+\frac{\Delta^2}{8\omega_rp\gamma\pi} .
\label{dpreg}
\ee

 Thus, we derived the formulas for the momentum diffusion
coefficient $D_p$ in the regimes of Hamiltonian chaos
(\ref{dpc}) with $\Lambda=1$ and Hamiltonian order (\ref{dpreg}) with $\Lambda=0$. In a general case with $0\leq\Lambda\leq1$, we will  assume a linear combination
\be
\begin{aligned}
D_p\simeq(1-\Lambda)D_p^{RN}+\Lambda D_p^{ch}\simeq\\\simeq\frac{\gamma}
{{12}}+\frac{\Delta^2}{8\omega_rp}\left(  \frac{  1-\Lambda }{ \gamma \pi  }+\frac{\Lambda}{\omega_rp}\right)\label{dprdpc}.
\end{aligned}
\ee
This function, shown by the solid line in the right upper panel in Fig.~\ref{fig7}, fits rather well
 exact numerical results.

 We would like to end this section with the proposal of
a simple experimental scheme to observe our main theoretical
and numerical result on an abrupt change in the character of atomic diffusion
in a laser standing wave under conditions corresponding to two different regimes of the underlying Hamiltonian
evolution, chaotic and regular ones. Let us consider a small
atomic cloud moving in one direction with
an average atomic momentum $\aver{p_c}$. Initial position and momentum
distribution are assumed to be a Gaussian with
the standard deviations $\sigma^2_x\equiv\aver{(x-\aver{x_c})^2}$ and $\sigma^2_p\equiv\aver{(p-\aver{p_c})^2}$.
The momentum diffusion coefficient is
\be
D_p=\frac{d(\sigma^2_p)}{2d\tau}.
\label{dtau}
\ee
 The temperature of gas and its rate of heating in Kelvins per second are
\begin{equation}
T\equiv\frac{2\aver{E_k}}{k_B}=\frac{\hbar^2k^2_f\sigma^2_p}{m_ak_B},
\quad\frac{dT}{dt}=\frac{2\hbar^2k^2_f\Omega D_p}{m_ak_B},\label{dt}
\end{equation}
where $E_k$ is a kinetic energy of atoms
 (in Joules) in the reference frame moving with the CM of the
 cloud. It follows from (\ref{dt}) that the rate of
heating is proportional to $D_p$ whose behavior is different in
the regimes of regular and chaotic underlying Hamiltonian dynamics.

The linear extent of the cloud in meters is $L_X\equiv2\sigma_x/k_f$.
On a comparatively short time scale, $\tau\ll |(\aver{p_c}/F)|$, and low temperatures $\sigma_p\ll|\aver{p_c}|$,
$D_p$ is approximately the same for all the atoms in the
cloud because the CM velocity could not change
significantly under the action of the friction force $F$ during
the observation time. Using the first equation in the set (\ref{mainsys}) and the Eq.(\ref{dtau}), we obtain
\be
 \sigma^2_x\simeq\sigma^2_x(0) + 
\frac{1}{2}\omega^2_r\sigma^2_p(0)\tau^2+\frac{2}{3}D_p\omega^2_r\tau^3.\label{extent}
\ee  
\begin{figure}
\includegraphics[width=0.47\textwidth,clip]{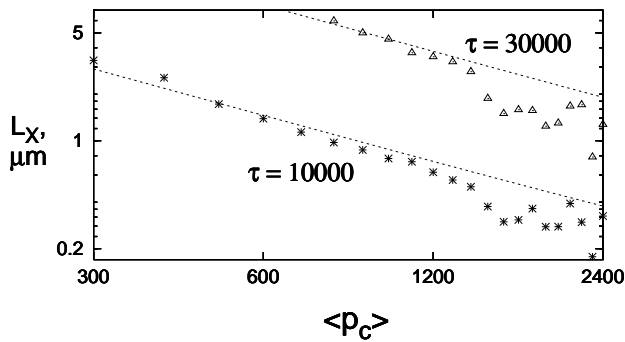}
\caption{A log-log dependence of the cloud's linear extent $L_X$ (in microns) 
on the average initial momentum $\aver{p_c}$ of atoms in a cloud 
at two moments of time. The analytic dashed lines 
were computed with the formula (\ref{extent}) with $D_p=D_p^{ch}$   valid in the chaotic regime 
on a short time scale. Note an abrupt change in the decay of $L_X(\aver{p_c})$
in the range of $\aver{p_c}\simeq 1200$ where a chaos-order transition takes place 
in the underlying Hamiltonian motion. $\Delta=-0.01$, $2\sigma_x(0)=0.5$, 
$2\sigma_p(0)=5$, wavelength $\lambda_f=2\pi/k_f=850$ nm.}
\label{fig8}
\end{figure}
We have computed $L_X$ with 
 that formula with $D_p$ given by (\ref{dpc}) and
compare the result with numerical simulation of Eqs.~(\ref{mainsys}) for a number
of atomic clouds with different initial values of $\aver{p_c}$.
In Fig.~\ref{fig8} the dependence $L_X(\aver{p_c})$ is plotted
for $\Delta=-0.01$ at two moments of time. The analytic
dashed lines fit well the exact numerical results
in the range $\aver{p_c}\lesssim1200$ where the underlying
Hamiltonian dynamics is chaotic (see the left column
in Fig.~\ref{fig7}). Note an abrupt change in the decay
of $L_X(\aver{p_c})$ beginning with those values
of $\aver{p_c}\simeq1200$ where the Hamiltonian dynamics
becomes more regular. Since the linear extent
of the atomic clouds changes abruptly at the
chaos-order border one may conclude that in real
experiments the value of $L_X$ should increase sharply with
those values of the average cloud momentum $\aver{p}$ for
which the underlying Hamiltonian evolution is chaotic.

\section{Conclusion}

Coherent evolution of the atomic
state in a near-resonant standing-wave laser field is interrupted
by SE events at random moments of times. The Hamiltonian
evolution between these events has been shown
previously (for a summary of Hamiltonian theory for cold atoms in a 1D optical lattice see Ref. \cite{pra07}) to be
regular or chaotic depending on
the values of the detuning $\Delta$ and the initial momentum $p_0$.
We stress that dynamical chaos may happen without any
noise and any modulation of the lattice
 parameters. It is a specific kind of
dynamical instability in the fundamental interaction
between the matter and radiation.

In reality Hamiltonian chaos is masked by random events of SE.
The behavior of spontaneously emitting atoms in the detuning
and momentum regimes where the underlying
Hamiltonian dynamics is chaotic may be called
{\it stochastic chaos}. We have specified and
quantified two regimes of the stochastic chaos, namely, random walking and dissipative ballistic transport. In the first
 regime, atoms wander in an optical lattice
 in a random way performing flights in both the directions with the PDFs strongly
depending on the detuning (see Figs. \ref{fig5}
and \ref{fig6}). In the ballistic regime, atoms
move in the same direction with momentum fluctuations
caused both by the Hamiltonian instability as well as
SE events. It has been shown in our numerical experiments
and confirmed analytically that the character of 
momentum diffusion changed abruptly in the regime
where the underlying Hamiltonian dynamics
 proved to be chaotic. A clear correlation between
the decay of the momentum diffusion coefficient $D_p$
and probability of Hamiltonian chaos $\Lambda$ has
been found (Fig.~\ref{fig7}). In order to observe
the manifestation of Hamiltonian chaos in real
experiments we proposed to measure a linear
extent of atomic clouds $L_X$ in a 1D optical lattice
and predicted a significant increase in $L_X$ for
the atomic clouds with $\Lambda\simeq1$.

In conclusion we would like to discuss some
possible applications of the theory developed and
the results obtained. A sensitive dependence of
statistical properties of dissipative chaotic
walking and ballistic transport on the values
of the detuning $\Delta$ provides a possibility to manipulate
atomic CM motion by changing $\Delta$. For example,
 one can increase the mean duration of
atomic flights in three orders of magnitude by
changing $\Delta$ only by thirty percents (see Fig.~\ref{fig6}).

Cold atoms in optical lattices is an ideal system to
study different phenomena in statistical physics.
Besides dynamical chaos,
the phenomena of stochastic resonance has been observed
in a near-resonant optical lattice \cite{SC02}. Another
phenomenon of considerable current interest is cold atom
ratchets \cite{ML99,SS03,JG04,GB07,GP07}.
A ratchet is a spatially periodic device which is able to produce
a directed transport of particles in the absence of a net bias
(i. e., when the time- and space-averaged forces are zero). In order to
realize the ratchet effect it is necessary to break time
or/and spatial symmetries which generate a countermoving partner to each trajectory \cite{FY00}. Different classes
of the ratchets have been experimentally realized with
cold atoms in optical lattices \cite{ML99,SS03,JG04}.
The interrelation of Hamiltonian chaos and SE noise, found in
this paper, provides an additional possibility to create and manipulate directed
transport of atoms in rigid optical lattices.

\section*{Appendix}

 We work in
the regime of small detunings
\be
|\Delta|\ll 1,
\label{delta}
\ee
 moderate mean atomic velocities
\be
\aver{|\omega_rp|} \sim\gamma/2\ll 1,
\label{omega}
\ee
and diffusive motion
\be
\tau\gg 2/\gamma.
\label{gamma}
\ee
 Due to (\ref{delta}), we may neglect
the last term in the potential energy (\ref{H}) and adopt the Hamiltonian solutions for (\ref{vz})
 between any two acts of SE.
The evolution of $u$ is described by the approximate solutions (\ref{udeterm}) for the regular Raman-Nath motion
 and (\ref{u_m}) for the chaotic motion. It follows from the condition of moderate atomic velocity (\ref{omega}) and solutions (\ref{udeterm}), (\ref{u_m}) that 	$|u|\ll 1$. In other words, $u$ cannot go far from zero between
 acts of SE after each of which $u=0$. Under the conditions (\ref{delta}) and $|u|\ll 1$, the mean kinetic atomic energy is much greater than the potential one and in this {\it weak Raman-Nath approximation}
\be
|u\sin x+\frac{\Delta z}{2}|\ll\frac{\omega_rp^2}{2},
\label{Hsim}
\ee
the solution (\ref{vz}) for $z$ is simplified
\begin{equation}
\begin{aligned}
z(\tau)\simeq\mp\sin\left(2\int\limits_0^\tau \cos  x d\tau'+
\chi_0\right).
\label{vz2}
\end{aligned}
\end{equation}
In the diffusion regime (\ref{gamma}) and in the weak Raman-Nath approximation (\ref{Hsim})
 the momentum fluctuations between two SE are small in comparison
with its fluctuations over the time scale of atomic transport. So, the atomic momentum just before SE is equal to its value at the node
\be
p\simeq p(\tau_j^-)\simeq p_{\rm node}.
\label{ptauj}
\ee
Now we can get simplified solutions for $u$. In the {\it exact Raman-Nath
condition}, $x=\omega_rp\tau$, we have from (\ref{udeterm}) the approximate deterministic map written in a two-step form
\begin{equation}
\begin{aligned}
u_m^{RN}\approx2\Delta 
 \sqrt\frac{\pi}{\omega_r p}\ v_0\cos\left(\frac{2}{\omega_r p}-\frac{\pi}{4}\right)+u_{m-2}^{RN}.\label{u2determ} 
\end{aligned}
\ee
 In the chaotic regime we have from $(\ref{u_m})$ the stochastic map
\begin{equation}
\begin{aligned}
\ &u_m^{ch}=|\Delta|\sqrt\frac{\pi}{\omega_r p}
\sin\phi_{m}+u_{m-1}^{ch}.
\end{aligned}
\label{u2_m}
\end{equation}

The maps derived enable us to estimate the values of $u(\tau_j^-)$ just before $j$th act of SE.
After the $(j-1)$th SE $u_0=u(\tau_{j-1}^+)=0$, $u(\tau_j^-)=u_M$ is
 an accumulated value of $u$ after passing $M$ nodes in the interval $\tau_{j-1}<\tau<\tau_j$. The
average number of node crossings can be estimated to be $\aver{M}\simeq 2|\omega_rp|/\gamma\pi$.
In the exact Raman-Nath limit, $u_M=M(u_m-u_{m-2})/2$, and the mean squared value in the regular regime
\be
\aver{u^2(\tau_j^-)}_{RN}=\aver{u^2_M}_{RN}\simeq
 \frac{4\Delta^2\omega_r p}{\pi\gamma^2}\ \aver{v^2_0}\cos^2\left(\frac{2}{\omega_r p}-\frac{\pi}{4}\right).\label{u2determ2} 
\ee
Beyond the strong Raman-Nath limit, $u(\tau_j^-)=u_M$ is a sum of $M$ random
numbers which are proportional to $\sin\phi_{m}$. From the probability
theory we get 
\be
\aver{u^2(\tau_j^-)}_{ch}=\aver{u^2_M}_{ch}\simeq\aver{M}\Delta^2\frac{\pi}{2\omega_r p}\simeq\frac{\Delta^2}{\gamma}.\label{u2_m2}
\ee

\end{document}